\documentclass[12pt]{article} 
\usepackage{graphicx}
\newcommand{\be}{\begin{equation}}
\newcommand{\ee}{\end{equation}}
\newcommand{\bear}{\begin{eqnarray}}
\newcommand{\eear}{\end{eqnarray}}
\newcommand{\ba}{\begin{array}}
\newcommand{\ea}{\end{array}}

\newcommand{\CL}{{\cal L}} 
\newcommand{\CN}{{\cal N}} 
 
\newcommand{\CG}{{\cal G}} 
 
\newcommand{\CO}{{\cal O}}

\begin{document}

\begin{titlepage}
\vfill
\begin{flushright}
{\normalsize KIAS-P06043}\\
{\normalsize hep-th/0612002}\\
\end{flushright}

\vfill
\begin{center}
{\Large\bf AdS/CFT with Tri-Sasakian Manifolds }

\vskip 0.3in

{ Ho-Ung Yee\footnote{\tt ho-ung.yee@kias.re.kr}}

\vskip 0.15in

 {\it Korea Institute for Advanced Study,} \\
{\it 207-43, Cheongryangri 2-dong, Dongdaemun-gu, Seoul 130-722,
Korea}
\\[0.3in]

{\normalsize  2006}

\end{center}

\vfill

\begin{abstract}
\normalsize\noindent We consider generic toric Tri-Sasakian 7-manifolds $X_7$ in the context of M-theory
on $AdS_4\times X_7$ and study their AdS/CFT correspondence to $\CN=3$ SCFT in 3D spacetime.
We obtain volumes of Tri-Sasakian manifolds and their supersymmetric 5-cycles via cohomological integration
technique, and use this to calculate conformal dimensions of baryonic operators in the SCFT side.
We also propose quiver-type gauge theories for UV description of the corresponding $\CN=3$ SCFT.
\end{abstract}

\vfill

\end{titlepage}
\setcounter{footnote}{0}

\baselineskip 18pt \pagebreak
\renewcommand{\thepage}{\arabic{page}}
\pagebreak

\section{Introduction }

AdS/CFT correspondence is a conjectured duality between a field theory living on
solitonic extended objects and a theory of gravity in the corresponding back-reacted space-time
near the extended objects \cite{Maldacena:1997re,Gubser:1998bc,Witten:1998qj}.
Its usefulness rests on the fact that the strong coupling regime of the field theory side
is mapped to a weakly interacting regime of the gravity theory, allowing us to calculate
non-trivial information about the strongly coupled field theory.
In addition to the original example of D3 branes in flat (9+1)-dimensional space-time,
corresponding to the duality between $\CN=4$ SYM in (3+1)-dimension and Type IIB string on $AdS_5\times S^5$,
there are many other examples with less number of supersymmetries as well as with
different space-time dimensions.
For example, when we consider D3 branes at the tip of a 6-dimensional cone,
we would have a supersymmetric 4D CFT at low energy with the number of supersymmetries determined by the
special holonomy of the cone.
As we consider supergravity back-reaction of the D3 branes, the resulting near horizon
geometry is $AdS_5\times X_5$, where $X_5$ is the constant radius section of the cone.
The AdS/CFT correspondence claims that the Type IIB string theory on this geometry
describes the corresponding 4D SCFT on D3 branes \cite{Acharya:1998db,Morrison:1998cs}.

A well-studied class of examples is the case when our
6-dimensional cone is locally Calabi-Yau, or Ricci-flat Kahler.
The constant radius section is then a Sasaki-Einstein manifold,
and the corresponding 4D SCFT is $\CN=1$. The simplest example of
these is provided by conifold, or its constant radius section
$T^{1,1}$, whose dual $\CN=1$ SCFT was first studied by Klebanov
and Witten \cite{Klebanov:1998hh}. In fact, this is the simplest
example of toric CY cones that can be obtained through Kahler
quotient from flat higher dimensional space. However, the Kahler
quotient only dictates and guarantees the Kahler class where
Ricci-flat metric belongs to, and the actual Ricci flat metric is
not given by the induced metric from the original flat space.
Finding the Ricci-flat metric is not an easy problem, and the
recently found $Y^{p,q}$ and $L^{p,q,r}$ are among the examples
\cite{Gauntlett:2004yd,Cvetic:2005ft}. It is also closely related
to the recent Z-minimization approach
\cite{Martelli:2005tp,Martelli:2006yb}. There also appeared
several interesting proposals for the corresponding $\CN=1$ SCFT's
in terms of quiver-type gauge theories
\cite{Martelli:2004wu,Benvenuti:2004dy,Benvenuti:2004wx,Hanany:2005hq,Benvenuti:2005ja,Franco:2005sm,Butti:2005sw}.
Predictions on $U(1)_R$ symmetry via a-maximization
\cite{Intriligator:2003jj} are found to match nicely to the
corresponding Reeb vector in the local CY cone via Z-minimization,
supporting these proposals in a convincing way
\cite{Bertolini:2004xf,Butti:2005vn,Lee:2006ru}.

In this paper, we consider another class of AdS/CFT examples in M-theory.
Specifically, we study 3D $\CN=3$ SCFT's that arise from M2 branes sitting at the tip
of 8-dimensional toric hyper-Kahler cones. The corresponding M-theory dual is $AdS_4\times X_7$
with $X_7$ being Tri-Sasakian manifolds.
Our discussion will include generic toric hyper-Kahler
cones obtained by arbitrary $U(1)^r$ hyper-Kahler quotients from flat spaces.
Although these spaces are well-known mathematically \cite{Boyer:1998sf,bgm2,Esch1,Bielawski},
the analysis in the context of AdS/CFT correspondence has not been done in full generality except
simple special cases \cite{Fre':1999xp,Fabbri:1999hw,Billo:2000zr,Gauntlett:2005jb,Lee:2006ys}.
Contrary to local Calabi-Yau cases in both Type IIB and M-theory, the structure
of hyper-Kahler manifold is sufficiently rigid such that the induced metric from the original
flat space is automatically hyper-Kahler that we would need.
This is essentially due to the rigidity of non-Abelian nature of $SU(2)_R$ structure of hyper-Kahler manifolds.
In the $\CN=3$ SCFT side, this is reflected to the fact that $SU(2)_R$ R-symmetry does not change
under RG-flows.
This is much help to us since R-charges of chiral primary operators encode information on their
conformal dimensions \cite{Minwalla:1997ka}. RG-rigidity may enable us to extract non-trivial information about
far IR physics from a simple UV description that is supposed to flow to our SCFT at IR.
This information on $\CN=3$ SCFT can then be compared to the results from its gravity dual description,
that is, M-theory on $AdS_4\times X_7$.

We will study M5 brane wrapping a supersymmetric cycle inside $X_7$ in $AdS_4\times X_7$.
(For a recent analysis for Type IIB case, see Ref.\cite{Butti:2006au}).
In view of $AdS_4$, this looks like a very  heavy point-like excitation.
In full second quantized
quantum theory, this will be described by an effective field in $AdS_4$ with a very large mass term.
Considering supersymmetry, there should actually be a super-multiplet arising from quantization of M5 world-volume theory.
Via a standard AdS/CFT dictionary, this super-multiplet with a heavy mass corresponds to a chiral
primary operator in $\CN=3$ SCFT whose conformal dimension is determined from the $AdS_4$ mass.
For a heavy mass compared to the curvature scale of $AdS_4$, the conformal dimension is proportional to the mass
of the super-multiplet in $AdS_4$, which is again proportional to the volume of the cycle $\Sigma_5$ that M5 brane
is wrapping.
The result of these calculations gives us
\be
\Delta = {\pi N\over 6}{{\rm Vol}(\Sigma_5)\over {\rm Vol}(X_7)}\quad,
\ee
where $\Delta$ is the conformal dimension of the corresponding chiral primary operator, and $N$
is the number of background M2 branes \cite{Gubser:1998fp}.
Therefore, from calculations on volumes of $\Sigma_5$ and $X_7$ we can extract $\Delta$.
This is then compared to predictions from a UV description of our $\CN=3$ SCFT in terms
of quiver-type gauge theory \cite{Fabbri:1999hw,Gauntlett:2005jb,Lee:2006ys}. We stress that this is meaningful due to RG-rigidity of $SU(2)_R$
R-symmetry that encodes information on conformal dimensions at IR  SCFT fixed point.

Though we will find a nice agreement later in this context of wrapping M5 brane,
this UV description is not completely satisfactory. Some of chiral primary spectrum
in the UV description is in fact shown to be absent in M-theory on $AdS_4\times X_7$ for the simplest
case of homogeneous Tri-Sasakian manifold $N(1,1)$ \cite{Fabbri:1999hw,Gauntlett:2005jb}.
This suggests that chiral primary spectrum may jump as we flow to the strong coupling regime at IR.
We don't see any tool to extract information about these phenomena.
This elimination of spectrum makes sense,
since we know that the number of degrees of freedom at the IR
fixed point scales mysteriously as $N^{3\over 2}$ while our UV description in terms of gauge theory has $N^2$
degrees of freedom. (For a recent account of $N^{3}$ scaling of M5 branes, see Ref.\cite{Lee:2006gq}.)

Recently, there appeared an alternative description of $\CN=2$ SCFT's arising from M2 branes
at the apex of Calabi-Yau cones, in terms of crystal lattice of M5 branes \cite{Lee:2006hw}.
Since $\CN=3$ belongs to $\CN=2$, it would be interesting to study further in that context.


\section{Quotient and Localization}

In this section, we discuss necessary technical gadgets to
calculate symplectic volume of toric tri-Sasakian manifolds.
These manifolds will arise as hyper-kahler quotients from higher dimensional
flat quaternion spaces.
In general, the quotient manifold that we are interested in is
a curved manifold, on which it is difficult to do explicit calculations.
A basic motivation of equivariant cohomology is to
develop a method that describes the usual cohomology of the symplectic quotient space
in terms of a language of the flat ambient space where calculations may
become much easier.

Perpendicular to this, there is a technique of localization.
It is often the case that the integration of our interest, like symplectic
volumes, has a fermionic nilpotent symmetry.
We introduce concepts of cohomology with it.
If there is a bosonic global symmetry,
we can use it to deform the fermionic nilpotent symmetry and its cohomology in a specific way,
parameterized by $\epsilon$'s.
To keep invariance under the defomed fermionic symmetry, the
integrand should also be modified. The defomed fermionic symmetry then
allows us to add a cohomologically trivial term in the integrand without affecting the result,
which contains, among other things, a positive definite purely bosonic term.
By taking the overall coefficient infinitely large,
the integration localizes to saddles points of the bosonic term, which
turn out to be nothing but the fixed points of the global symmetry we started with.

For compact manifolds, we may take a continuous limit of turning off the deformation
parameter $\epsilon$'s to get the results for the original problem.
For non-compact manifolds, the deformation often provides a regularization and
we may get other information from the results of the deformed integral.
The volume of tri-Sasakian manifolds that we will discuss belongs to the latter case.

In many cases, the symmetry we use for quotient is compatible with the symmetry of
the quotient space for $\epsilon$-deformation. In other words, the symmetry
for $\epsilon$-deformation in the quotient space
is actually a symmetry in the ambient space, too.
We are then able to describe $\epsilon$-deformation and the resulting localization
for the quotient space in a language of equivariant cohomology in flat ambient space.
This gives us a powerful handle over easy calculations in flat spaces
of seemingly difficult integrals in curved quotient spaces.

The techniques presented in this section have been established in Ref.\cite{Lee:2006ys,Witten:1982im,Moore:1997dj,park}, and
for completeness we will expound them in more explicit detail.
Readers familiar to it may skip this section.

{   \it A supermanifold $T[1]X$}

Integrals of our interest are typically those of differential forms,
and it is possible to rewrite them in a way that looks like a supersymmetric
path integral. Though this is not a strictly necessary formulation,
it may give us a comfortable understanding of some mathematical results in
physics terms.

Given a bosonic manifold $X$ with a coordinate $\{x^\mu\}$, a tangent vector $V$
is canonically written as $V=V^\mu {\partial \over \partial x^\mu}$. We can think of
$\{x^\mu, V^\mu\}$ as a canonical coordinate system of the tangent bundle $TX$ associated
to a coordinate $\{x^\mu\}$.
The supermanifold $T[1]X$ is obtained by
replacing the bosonic coordinates $\{V^\mu\}$ with fermionic ones $\{\psi^\mu\}$ to which
we assign a degree number 1, hence explaining the notation.
Functions on $T[1]X$ will be expanded as
\be
f(x,\psi)=f^{(0)}(x)+f^{(1)}_\mu(x) \psi^\mu+{1\over 2!}f^{(2)}_{\mu\nu}(x)\psi^\mu \psi^\nu+\cdots \quad,
\ee
up to the dimension $n$ of the manifold $X$,
and the space of functions on $T[1]X$ is easily identified as $\Omega^*(X)$,
the space of differential forms on $X$.
An integration over $T[1]X$ of a function $f(x,\psi)$ is
\bear
\int_{T[1]X} [dx^\mu] [d\psi^\mu] \,\,f(x,\psi)&=&\int_{T[1]X}[dx^\mu] [d\psi^\mu]\,\,
{1\over n!}f^{(n)}_{\mu_1 \mu_2 \cdots \mu_n}(x)\psi^{\mu_1} \psi^{\mu_2} \cdots \psi^{\mu_n}\nonumber\\
&=&\int_X [dx^\mu] \,\,{1\over n!}f^{(n)}_{\mu_1 \mu_2 \cdots \mu_n}(x)\epsilon^{\mu_1 \mu_2 \cdots
\mu_n}= \int_X \,\,f^{(n)}\quad,
\eear
which is the usual integration of top differential form on $X$. Note that the top form in $f(x,\psi)$
is picked up automatically by the fermionic integration over $\psi^{\mu}$.
In fact, the measure $[dx^\mu][d\psi^\mu]$ is invariant under coordinate transformations.
Since $\psi^\mu$ transforms as a vector under a coordinate change $x^\mu\rightarrow \tilde{x}^\mu$,
we have $\tilde\psi^\nu=\psi^\mu\left(\partial\tilde x^\nu\over \partial x^\mu\right)$
and $[d\tilde\psi]=\det\left(\partial\tilde x^\nu\over \partial x^\mu\right)^{-1}[d\psi]$,
which cancels the bosonic Jacobian.

The supermanifold $T[1]X$ has a nilpotent fermionic symmetry,
\bear
Q\,x^\mu &=& \psi^\mu\quad, \nonumber\\
Q\,\psi^\mu &=& 0\quad,
\eear
with $Q^2=0$. An inspection of its action to functions on $T[1]X$
shows that it is nothing but the usual de Rham differential acting on $\Omega^*(X)$, $Q \simeq d$.
We define observables of our supersymmetric theory as $Q$-cohomology classes, and
correlation functions of them are integrals on $T[1]X$, which are identical to
intersection integrals of $X$.

The measure on $T[1]X$ is invariant under $Q$, and from this we can derive that
\be
\int_{T[1]X}\,\,Q\Lambda(x,\psi) =0 \quad,\label{exact}
\ee
for any $\Lambda$. This is a compact form of the Stokes theorem in the following sense.
For a cycle $Y$ in $X$, there corresponds a Poincare dual form $\delta(Y)$ in $Q$-cohomology,
such that
\be
\int_{T[1]Y}\,f =\int_{T[1]X} f\cdot \delta(Y)\quad,
\ee
for any $f$. Let's define the boundary of $Y$ such that its Poincare dual is $Q\delta(Y)$, that is,
$\delta(\partial Y)= Q\delta(Y)$. It follows that
\be
0=\int_{T[1]X}\,Q(f \cdot\delta(Y))=\int_{T[1]X}\,Qf\cdot\delta(Y)\pm \int_{T[1]X}\,f\cdot Q\delta(Y)
=\int_{T[1]Y} \, Qf \pm\int_{T[1]\partial Y}\, f\quad,
\ee
where $\pm$ depends on the degree numbers.

We are interested in calculating volumes of symplectic(Kahler) manifolds $(X,\omega)$ of dimension $2n$
with a non-degenerate symplectic 2-form $\omega={1\over 2}\omega_{\mu\nu}\psi^\mu\psi^\nu$.
Since the volume element is simply ${1\over n!}\omega^n$, it can be written as
\be
{\rm Vol}(X)=\int_{T[1]X}\,e^\omega = \int_{T[1]X}\,e^S\quad,\label{volume}
\ee
with an action $S=\omega$ which is $Q$-symmetric, $QS=Q\omega=0$, from $d\omega=0$.
The above looks like a (0+0)-dimensional supersymmetric partition function.
From (\ref{exact}), we are free to add $Q$-exact terms to the action, $S\rightarrow S+Q\CO$,
without affecting the result.
However, since $Q\CO$ contains at least one $\psi^\mu$, there is no purely bosonic term
we can add  to facilitate localization argument in this case.
This is one motivation to introduce $\epsilon$-deformation using global symmetries of $\omega$.

{\it  Hamiltonian Flow}

Symmetries of a symplectic manifold $(X,\omega)$ is locally isomorphic
to the space of real functions on $X$ up to a constant function.
More explicitly, for any function $H$ on $X$ we can define a vector field $V$ from $QH=i_V \omega$,
where $i_V$ is the contraction by the vector $V$,
\be
i_V=V^\mu {\partial\over \partial \psi^\mu}\quad.\label{iv}
\ee
In components, we have ${\partial H\over \partial x^\nu}=V^\mu \omega_{\mu\nu}$,
and $H$ determines $V$ because $\omega$ is non-degenerate and we can invert it.
From $\CL_V=\{i_V,Q\}=i_V Q+Q i_V$, where $\CL_V$ is the Lie derivative by $V$,
and $Q\omega=0$, we have
\be
0=Q^2 H =Q i_V\omega =\{i_V,Q\}\omega = \CL_V \omega\quad,
\ee
saying that $V$ generates a symmetry flow of $\omega$. Conversely, given a symmetry vector $V$,
\be
0=\CL_V \omega =\{i_V,Q\}\omega =Q i_V\omega \quad,
\ee
and we can always find $H$ with $QH=i_V \omega$ at least locally up to an additive constant.

The space of symmetries of $\omega$ is closed under Lie bracket, and the corresponding
operation in the space of functions turns out to be Poisson bracket with respect to $\omega$.
In other words, letting $V_f$ and $V_g$ be symmetries obtained from functions $f$ and $g$,
their Lie bracket $[V_f , V_g]$ is equal to $V_{\{f,g\}}$, a symmetry from the Poisson
bracket $\{f,g\}$ with respect to $\omega$. Therefore, the correspondence is an algebra isomorphism.
The Jacobi identity for Poisson bracket is a simple consequence of that for Lie bracket of vector fields.

{\it Deformation via Global Symmetries and Localization}

In (\ref{volume}), the symplectic volume is written in a $Q$-symmetric way, but
a freedom of adding $Q$-exact terms to the action doesn't help much for its calculation.
However, by deforming $Q$-symmetry using  global symmetries and also the action accordingly,
there is a way to use this
freedom to reduce our integration into a localized sum over discrete points.
After an easy calculation of the deformed integral via localization, one may turn off the deformation
parameter $\epsilon$ to get the answer for the original integral.

Suppose we have a well-defined symmetry flow $V=V^\mu {\partial\over \partial x^\mu}$,
and the corresponding function $H$ with $QH=i_V \omega$. This allows us to deform $Q$ into
\bear
Q_{\epsilon} \,x^\mu &=& \psi^\mu \quad,\nonumber\\
Q_{\epsilon}\, \psi^\mu &=& \epsilon V^\mu(x)\quad,
\eear
with $\epsilon$ a real number. This can also be written as $Q_{\epsilon}=Q+\epsilon i_V$,
with $i_V$ in (\ref{iv}). From $Q^2=(i_V)^2=0$, it is readily seen that $Q_{\epsilon}^2 =\epsilon \CL_V$,
which implies $Q_{\epsilon}$ is nilpotent only in the subspace of $V$-invariant functions on $T[1]X$.
We henceforth must restrict to this $V$-invariant subspace when we discuss adding
$Q_\epsilon$-exact terms later on.

The original action $S=\omega$ is not $Q_\epsilon$-invariant,
\be
Q_\epsilon S=Q\omega+\epsilon i_V\omega = \epsilon QH= Q_\epsilon (\epsilon H)\quad.
\ee
Instead, the deformed action $S_\epsilon=S-\epsilon H$ is $Q_\epsilon$-invariant.
Note that $S_\epsilon$ is also $V$-invariant from $(i_V)^2=0$. Therefore,
one considers the deformed volume which is $Q_\epsilon$-invariant,
\be
{\rm Vol}_\epsilon(X)=\int_{T[1]X}\,\,e^{S_\epsilon}=\int_{T[1]X}\,\,e^{S-\epsilon H}\quad,
\ee
and an arbitrary $Q_\epsilon$-exact term $Q_\epsilon \CO$ may be added to the action
without changing the result
as long as $\CO$ is $V$-invariant, $\CL_V \CO=0$. One such $Q_\epsilon\CO$ of our interest is
\bear
-t Q_\epsilon \left(g_{\mu\nu}\psi^\mu V^\nu\right)&=&-t\left(\epsilon g_{\mu\nu} V^\mu V^\nu
+\partial_\eta(g_{\mu\nu}V^\nu)\psi^\eta \psi^\mu\right)
\eear
with a $V$-invariant positive definite metric $g_{\mu\nu}$ on $X$.
We can always find it by averaging any metric along $V$-flows.
Note that we have an expected bosonic term, $-t\epsilon g_{\mu\nu}V^\mu V^\nu$
which is negative definite, and
taking $t\rightarrow +\infty$ limit reduces the whole integral to a saddle-point approximation
around fixed points of $V$,
which becomes strictly exact. Note that vanishing points of $V=0$ is nothing but the extreme points of $H$.
Performing quadratic expansion around a fixed point $x^\mu=0$,
\be
V^\mu = V^\mu_\alpha x^\alpha+\CO(x^2)\quad, \quad
g_{\mu\nu}=g^{(0)}_{\mu\nu}+\CO(x)\quad,
\ee
and calculating Gaussian integration, we have a formula by Duistermaat-Heckmann,
\be
{\rm Vol}_\epsilon(X)=\sum_{{\rm fixed \,points} \,p}(-1)^{\#_p}
e^{-\epsilon H(p)}\left(2\pi \over \epsilon\right)^n\left( \det(g^{(0)}_{[ \mu\hat\nu}V^\nu_{\eta ]})
\over \det(g^{(0)}_{\mu\nu}V^\mu_\alpha V^\nu_\beta)\right)^{1\over 2}\quad.
\ee


{\it Symplectic Quotient and Equivariant Cohomology}

In general, even localization calculation on our space $(X,\omega)$ is
not easy if it is a curved manifold.
If it can arise as a symplectic quotient from a larger flat space $(M,\omega)$,
the language of equivariant cohomology provides a method
for calculating things for $X$ in the ambient flat space $M$  more easily.
Moreover, in the case where a global symmetry of $X$ is actually a symmetry of $M$,
the $\epsilon$-deformation and the localization on $X$ can
also be described on $M$ equivariantly. Since localization probes only the tangent
space of the geometry, the relation between $X$ and $M$ becomes a linear one around
the fixed points, and the calculation on $M$ is  a tractable one.

For simplicity we will discuss $U(1)$ quotient only, although  generalizations to $U(1)^n $
as well as non-abelian groups are straightforward.
Starting from an ambient space $(M,\omega)$, whether flat or not, with a $U(1)$-symmetry flow
$V$, its symplectic quotient $M//U(1)$ is defined as follows.
From $\CL_V\omega =0$, there exists a function $\mu(x)$ with $Q\mu=i_V\omega$, which is called
a moment map. Note that we have a freedom of adding constant function to $\mu(x)$, and
it actually parameterizes a family of $M//U(1)$ that we define with $\mu(x)$.
Since $V^\mu {\partial \mu(x)\over \partial x^\mu}=i_V(Q\mu)=(i_V)^2\omega=0$,
the level surface $\mu^{-1}(0)$ is invariant under $V$-flow, in other words, $V$ is
a well-defined $U(1)$-flow on $\mu^{-1}(0)$. $M//U(1)$ is then defined as the usual
quotient of $\mu^{-1}(0)$ by $U(1)$, $M//U(1)=\mu^{-1}(0)/U(1)$.

To analyze the situation more clearly, we introduce a coordinate system
$\{x^i,x^v,x^n\}$, $i=1,\cdots,(dim M-2)$, of $M$ around $\mu^{-1}(0)$ such that
$\{x^i\}$ parameterize $\mu^{-1}(0)/U(1)$, and $V={\partial\over \partial x^v}$,
that is, $x^v$ is the Gauss coordinate of the $V$-flow. $x^n$ parameterizes the normal direction
to $\mu^{-1}(0)$.
The equation $Q\mu=i_V \omega$ in components reads as
\be
\omega_{vi}={\partial\mu\over \partial x^i}\quad,\quad \omega_{vn}={\partial\mu\over \partial x^n}\quad,
\label{comp}
\ee
and because $\mu=0$ on $\mu^{-1}(0)$, its derivative with respect to $x^i$ is also zero,
hence $\omega_{vi}=0$ on $\mu^{-1}(0)$.
Then the $vij$-component of $Q\omega=0$ gives us
\be
\partial_v\omega_{ij}=\partial_i\omega_{vj}-\partial_j\omega_{vi}=0\quad,
\ee
and we see that $\omega_{ij}$ is $V$-invariant on $\mu^{-1}(0)$.
Therefore, ${1\over 2}\omega_{ij}\psi^i\psi^j$ is a well-defined symplectic form on $M//U(1)=\mu^{-1}(0)/U(1)$
with its coordinates $\{x^i\}$, and this defines the symplectic quotient $(M//U(1),\omega)$.
Its symplectic volume is then
\be
{\rm Vol}(M//U(1))=\int_{T[1]M//U(1)}[dx^i][d\psi^i]\,\,e^{{1\over 2}\omega_{ij}\psi^i \psi^j}\quad.
\label{quotientvolume}
\ee

Our basic objective is to re-express (\ref{quotientvolume}) in a language of the ambient space $M$.
Since everything is independent of $x^v$, we can simply extend the bosonic integration to
include $\int dx^v$ and divide by ${\rm Vol}(U(1))=\int dx^v$. The integration is now over $\mu^{-1}(0)$
and we have
\bear
{\rm Vol}(M//U(1))&=&{1\over {\rm Vol}(U(1))}\int_{\mu^{-1}(0)}[dx^v][dx^i][d\psi^i]
\,\, e^{{1\over 2}\omega_{ij}\psi^i \psi^j}\\&=&
{1\over {\rm Vol}(U(1))}\int_{M}[dx^v][dx^n][dx^i][d\psi^i]
\,\, e^{{1\over 2}\omega_{ij}\psi^i \psi^j} \delta\left(\mu(x)\right)
\left(\partial\mu(x)\over \partial x^n\right)\quad,\nonumber
\eear
where we have extended the bosonic integration over the whole $M$ by introducing $\delta$-function
of $\mu(x)$ with an appropriate Jacobian factor for $x^n$ integration.
With an auxiliary variable $\phi$ to write $\delta(\mu(x))={1\over (2\pi)}\int[d\phi] e^{i\phi \mu(x)}$,
and also introducing $\psi^v$ and $\psi^n$ to write
\be
\left(\partial\mu\over \partial x^n\right)=\int[d\psi^v][d\psi^n]\,e^{
\left(\partial\mu\over \partial x^n\right) \psi^v \psi^n }=
\int[d\psi^v][d\psi^n]\,e^{
\omega_{vn} \psi^v \psi^n}\quad,
\ee
where we have used (\ref{comp}), we arrive at
\bear
{\rm Vol}(M//U(1))&=&
{1\over (2\pi) {\rm Vol}(U(1))}\int_{M}[d\phi][dx^v][dx^n][dx^i][d\psi^i][d\psi^v][d\psi^n]
\,\, e^{{1\over 2}\omega_{ij}\psi^i \psi^j+\omega_{vn}\psi^v\psi^n +i\phi \mu(x)}\nonumber\\
&=&{1\over (2\pi) {\rm Vol}(U(1))}\int_{T[1]M\otimes\phi}[d\phi][dx][d\psi]\,\,
e^{{1\over 2}\omega_{\mu\nu}\psi^\mu\psi^\nu+i\phi\mu(x)}\nonumber\\&=&
{1\over (2\pi) {\rm Vol}(U(1))}\int_{T[1]M\otimes\phi}\,e^S\quad,\label{integral}
\eear
where we have a complete $\omega={1\over 2}\omega_{\mu\nu}\psi^\mu\psi^\nu$ and the measure on $T[1]M$
in the ambient space $M$. Note that there is no contribution from $\omega_{in}\psi^i\psi^n$ in the fermionic
integration.
Though we have used a specific coordinate system to arrive at the above,
we see that the end result is coordinate invariant.

Our integration (\ref{integral}) is now over a supermanifold $T[1]M\otimes \phi$ and
our action $S=\omega+i\phi\mu(x)$ is a function on $T[1]M\otimes\phi$.
We find that $S$ is invariant under the following fermionic symmetry acting on $T[1]M\otimes\phi$,
\bear
\widetilde Q x^\mu &=& \psi^\mu\quad,\nonumber\\
\widetilde Q \psi^\mu &=& -i \phi V^\mu(x)\quad,\nonumber\\
\widetilde Q\phi &=& 0\quad,
\eear
with $\widetilde Q^2=-i\phi \CL_V\simeq 0$ on the subspace of $V$-invariant functions.
We can also write it as $\widetilde Q = Q-i\phi i_V$, with $Q$ the usual de Rham differential.

What we have done is to replace the symplectic form in the quotient space $T[1]M//U(1)$ with a
$V$-invariant $\widetilde Q$-closed object $S=\omega+i\phi\mu(x)$ in the ambient space $T[1]M\otimes\phi$.
Up to a numerical factor in front, we see that the original symplectic volume integral
over $T[1]M//U(1)$ is equal to the integral of the corresponding  $\widetilde Q$-closed form $e^S$
over the ambient space $T[1]M\otimes\phi$.
The claim of equivariant cohomology is a more general statement of the above relation
between $T[1]M//U(1)$ and $T[1]M\otimes\phi$.
The usual $Q$-cohomology, or de Rham cohomology, of functions on $T[1]M//U(1)$
is identical to the equivariant $\widetilde Q$-cohomology of $V$-invariant
functions on $T[1]M\otimes\phi$,
and their correlation functions also match with each other.
This provides an easy way of calculating things about the quotient space in the flat ambient space.

{\it Equivariant $\epsilon$-deformation}

Suppose that there is an additional $U(1)$ global symmetry on $M$ generated by $R$
with $QH=i_R\omega$, which is compatible with $V$,
the symmetry by which we perform the  quotient.
This means that $[V,R]=0$ and $i_RQ\mu=i_R i_V\omega=-i_Vi_R\omega =-i_V QH =0$,
that is, $\mu(x)$ is invariant under $R$-flow and $R$ is well-defined on $\mu^{-1}(0)$.
Under this assumption, since $R$ is a well-defined vector field on $\mu^{-1}(0)$,
\be
R=R^i {\partial\over\partial x^i}+R^v {\partial\over \partial x^v}\quad,
\ee
and from $V={\partial\over \partial x^v}$, the component equation of $[V,R]=0$ reads as
\be
\left(\partial R^i\over \partial x^v\right){\partial\over \partial x^i} +
\left(\partial R^v\over \partial x^v\right){\partial\over \partial x^v}=0\quad,
\ee
and we see that $R^i$ is $V$-invariant on $\mu^{-1}(0)$ and defines a vector field
$R^i {\partial\over\partial x^i}$ on $M//U(1)$.
Moreover, $i$-component of the equation $i_R\omega=QH$ on $\mu^{-1}(0)$ is
\be
R^j\omega_{ji}+R^v\omega_{vi}={\partial H\over \partial x^i}\quad,
\ee
and from $\omega_{vi}=0$ on $\mu^{-1}(0)$ as before,
we have $R^j\omega_{ji}={\partial H\over \partial x^i}$, which is a simple statement that
$i_R\omega=QH$ still holds on the quotient space with the same function $H$, and $R$
is also a symmetry on $M//U(1)$.

We are then able to perform $\epsilon$-deformation with a symmetry $R$
in the quotient space $M//U(1)$ by simply replacing $\omega$ with $\omega-\epsilon H$ as before,
and the usual $Q$ in $T[1]M//U(1)$ is now replaced by $Q_\epsilon=Q+\epsilon i_R$.
Using equivariant cohomology, we next try to reformulate these in terms of the ambient space
$T[1]M\otimes\phi$. The fermionic symmetry $\widetilde Q$ in the ambient space will be
modified to $\widetilde Q_\epsilon=\widetilde Q+\epsilon i_R=Q-i\phi i_V+\epsilon i_R$, and
the $V$-invariant $\widetilde Q_\epsilon$-closed object that corresponds to
the deformed action will naturally be $\omega+i\phi\mu(x)-\epsilon H$ on $T[1]M\otimes\phi$.
We also restrict to both $V$- and $R$- invariant subspace to ensure nilpotency of $\widetilde Q_\epsilon^2=0$.
Therefore, we arrive at
\be
{\rm Vol}_\epsilon(M//U(1))={1\over (2\pi){\rm Vol}(U(1))}\int_{T[1]M\otimes\phi}
\,\,e^{\omega +i\phi\mu(x)-\epsilon H(x)}\quad.\label{kahler}
\ee
We may add an arbitrary $V,R$-invariant $\widetilde Q_\epsilon$-exact term
to the action to facilitate localization in $T[1]M\otimes\phi$.
The result usually boils down to contour integrations over $\phi$.


{\it Extension to Hyperkahler Quotient}

We can generalize the previous discussions to hyperkahler quotients,
and the idea is similar in spite of a few technical complications.
A hyperkahler manifold $M$ has three kahler forms $\vec\omega$, and its
symplectic volume is simply defined in terms of one of them, say, $\omega^3=\omega$.
Once we pick up one kahler form $\omega$,
its $\epsilon$-deformation by global symmetries of $\omega$ and localization
are same as in the kahker manifolds. It is not important whether
the other $\omega^1$ and $\omega^2$ may or may not
be invariant under the global symmetry.

The difference arises in the quotient because hyperkahler quotient
is something more than kahler quotient. Suppose there is a
$V$-flow which is a symmetry of the three kahler forms $\vec\omega$, such that
$i_V\vec\omega=Q\vec\mu(x)$ with three moment maps $\vec\mu(x)$.
As before $V$-flow leaves invariant the codimension 3
submanifold $\vec\mu^{-1}(0)$ and the hyperkahler quotient is defined as
$M////U(1)=\vec\mu^{-1}(0)/U(1)$.
Introducing a local coordinate system $\{x^i,x^v,x^n\}$ of $M$ around $\vec\mu^{-1}(0)$,
where $x^i$ parametrize $\vec\mu^{-1}(0)/U(1)$, $V={\partial\over\partial x^v}$,
and $x^n$, $n=1,2,3$, are three coordinates normal to $\vec\mu^{-1}(0)$,
it is easily verified in the exactly same manner as before
that $\vec\omega_{vi}=0$, $\vec\omega_{vn}=\partial_n \vec\mu$,
and $\vec\omega_{ij}$ is $V$-invariant on $\vec\mu^{-1}(0)$.
The quotient space $M////U(1)$ then naturally inherits $\vec\omega_{ij}$ as its
tri-holomorphic kahler forms of hyperkahler structure.

After picking up $\omega^3=\omega$ to define the symplectic volume, it is written as
\be
{\rm Vol}(M////U(1))=\int_{T[1]M////U(1)}\,\,e^{{1\over 2}\omega_{ij}\psi^i\psi^j}\quad,
\label{hyper}
\ee
and to rewrite the above in a language of the ambient space $M$ as before, we have
\bear
\int_{T[1]M////U(1)}&=&{1\over {\rm Vol}(U(1))}\int_{\vec\mu^{-1}(0)}[dx^v][dx^i][d\psi^i]\nonumber\\
&=&{1\over {\rm Vol}(U(1))}\int_{M}[dx^v][dx^n][dx^i][d\psi^i]\prod_{a=1}^3\delta(\mu^a(x))
\det\left(\partial\mu^a(x)\over\partial x^n\right)\nonumber\\
&=&{1\over (2\pi)^3{\rm Vol}(U(1))}\int_{M}[d\vec\phi][d\vec\chi][dx^\mu][d\psi^i][d\psi^n]
\,\,e^{i\vec\phi\cdot\vec\mu(x)+\vec\chi\cdot\left(\partial_n\vec\mu\right)\psi^n}\quad,
\eear
where we introduce bosonic auxiliary variables $\vec\phi$ and fermionic $\vec\chi$ as well as $\psi^n$.
Using $\partial_n\mu^3=\omega^3_{vn}=\omega_{vn}$, and $\omega_{vi}=0$, as
well as calling $\chi_3=\psi^v$, we
have $\chi_3(\partial_n \mu^3)\psi^n =\psi^v\omega_{vn}\psi^n= \psi^v\omega_{v\mu}\psi^\mu$,
where $\mu$ runs now over all coordinates of $M$.
Similarly, $\chi_a(\partial_n\mu^a)\psi^n=\chi_a(\partial_\mu\mu^a)\psi^\mu=\chi_a Q\mu^a$ for $a=1,2$.
Inserting these to (\ref{hyper}), we obtain
\bear
{\rm Vol}(M////U(1))={1\over (2\pi)^3{\rm Vol}(U(1))}
\int_{T[1]M\otimes\vec\phi\otimes\chi_a}\,\,e^{\omega+i\vec\phi\cdot\vec\mu+\chi_a Q\mu^a}\quad,
\eear
where $a=1,2$ and in completing $\omega={1\over 2}\omega_{\mu\nu}\psi^\mu\psi^\nu$ in the action,
we have used the fact that the missing piece $\omega_{in}\psi^i\psi^n+{1\over 2}\omega_{mn}\psi^m\psi^n$
can be absorbed by shifting $\psi^i$ and $\chi_a$ variables.
Note that the end result is a coordinate-invariant expression.
Therefore, we see that the symplectic volume integral in the quotient space
can be written an an integral in the ambient supermanifold $T[1]M\otimes\vec\phi\otimes\chi_a$.

The action $S=\omega+i\vec\phi\cdot\vec\mu(x)+\chi_a Q\mu^a (x)$ is easily seen to be invariant
under a fermionic symmetry on $T[1]M\otimes\vec\phi\otimes\chi_a$,
\bear
\widetilde Q x^\mu &=& \psi^\mu\quad,\nonumber\\
\widetilde Q \psi^\mu &=& -i\phi_3 V^\mu(x)\quad,\nonumber\\
\widetilde Q \vec\phi &=& 0\quad,\nonumber\\
\widetilde Q \chi_a&=& -i \phi_a\quad, a=1,2\quad,\label{tildeQ}
\eear
which is also written as $\widetilde Q=Q-i\phi_3 i_V-i\phi_a{\partial\over\partial \chi_a}$,
with $\widetilde Q^2 = -i\phi_3 \CL_V \simeq 0$ in the subspace of $V$-invariant functions.
The natural expectation of equivariant cohomology for hyperkahler quotient
would be that the usual $Q$-cohomology on $T[1]M////U(1)$ is identical to the $\widetilde Q$-cohomology
on $V$-invariant functions of $T[1]M\otimes\vec\phi\otimes \chi_a$.

Now we turn to the question of equivariant description of an $\epsilon$-deformation in
the quotient space in terms of the ambient space $T[1]M\otimes\vec\phi\otimes\chi_a$.
Contrary to the previous kahler case, we can have two distinct situations.
The first case is where we have an additional symmetry $R$ in $M$ compatible with $V$, which
preserves all three kahler forms $\vec\omega$. In this case, the modification of $\widetilde Q$
is simply $\widetilde Q_\epsilon=\widetilde Q+\epsilon i_R$ with the modification of the
action $S_\epsilon=S-\epsilon H$ with $i_R \omega =QH$. The structure is essentially
identical to that for the kahler case. The more interesting case is where
the symmetry $R$ preserves only $\omega^3=\omega$ and it acts as an $U(1)_R$ rotation for
the other $\omega^1$ and $\omega^2$, that is, $\CL_R(\omega^1-i\omega^2)=2i(\omega^1-i\omega^2)$,
$\CL_R(\mu^1-i\mu^2)=R^\mu\partial_\mu(\mu^1-i\mu^2)=2i(\mu^1-i\mu^2)$ and $i_R\omega=QH$.
In this case, the deformed action $S_\epsilon=S-\epsilon H$ is invariant under
the symmetry
\bear
\widetilde Q_\epsilon x^\mu &=& \psi^\mu\quad,\nonumber\\
\widetilde Q_\epsilon \psi^\mu &=& -i\phi_3 V^\mu(x)+\epsilon R^\mu(x)\quad,\nonumber\\
\widetilde Q_\epsilon \phi_3 &=& 0\quad,\nonumber\\
\widetilde Q_\epsilon \phi_1 &=& 2 i \epsilon \chi_2\quad,\nonumber\\
\widetilde Q_\epsilon \phi_2 &=& -2 i \epsilon \chi_1\quad,\nonumber\\
\widetilde Q_\epsilon \chi_a&=& -i \phi_a\quad, a=1,2\quad,\label{deformedQ}
\eear
with $\widetilde Q_\epsilon ^2=-i\phi_3 \CL_V +\epsilon\CL_R$, where $\CL_R$ acts on $\phi_a$ and $\chi_a$
as
\be
\CL_R(\phi_1-i\phi_2)=2i(\phi_1-i\phi_2)\quad,\quad \CL_R(\chi_1-i\chi_2)=2i(\chi_1-i\chi_2)\quad.
\ee
Then, we have to restrict to $R$-invariant subspace of $T[1]M\otimes\vec\phi\otimes\chi_a$,
taking into account of the $R$-action on $\phi_a$ and $\chi_a$, when we discuss adding
$\widetilde Q_\epsilon$-exact terms to the action.

An interesting point is that we can always have one such term
\be
-i t \widetilde Q_\epsilon (\chi_a \phi_a)=-t\phi_a \phi_a-(4\epsilon t) \chi_1\chi_2\quad,
\ee
and by taking $t\rightarrow +\infty$ limit, $\phi_a$, $\chi_a$ integration is
dominated by this term and we can simply integrate them out leaving
\be
\int [d\phi_1][d\phi_2][d\chi_1][d\chi_2] \,\,e^{-t\phi_a\phi_a -(4\epsilon t) \chi_1\chi_2 }
= {\pi\over t}\cdot (4\epsilon t) = 4\pi\epsilon\quad.
\ee
Performing this, our $\epsilon$-deformed symplectic volume ends up to
\be
{\rm Vol}_\epsilon(M////U(1))={4\pi\epsilon \over (2\pi)^3{\rm Vol}(U(1))}
\int_{T[1]M\otimes\phi_3}\,\,e^{\omega+i\phi_3\mu^3(x)-\epsilon H}\quad,\label{regvolume}
\ee
which looks just like that of the kahler case (\ref{kahler}).
This formula will be our starting point in the next section.


\section{Volumes of Toric Tri-Sasakian Manifolds and Their
Supersymmetric Cycles}

A simple definition of Tri-Sasakian manifold is that
its metric cone is hyper-Kahler. In other words, a $(4n-1)$-dimensional
manifold $X_{4n-1}$ is Tri-Sasakian when its cone with the metric
\be
ds^2_{4n}=dr^2+r^2 ds^2_{4n-1}\quad,
\ee
is a $(4n)$-dimensional hyper-Kahler cone, where $ds^2_{4n-1}$ is normalized to satisfy $R_{ij}=(4n-2)g_{ij}$.
Starting from flat quaternion spaces, we can construct non-trivial hyper-Kahler cones by the process
of hyper-Kahler quotient, and we subsequently obtain Tri-Sasakian
manifolds as the constant radius section of the cones \cite{Boyer:1998sf,bgm2,Esch1,Bielawski}.
As we will consider only abelian symmetries $U(1)^r$ of  flat quaternion
spaces in performing hyper-Kahler quotient, the resulting hyper-Kahler
cones or Tri-Sasakian manifolds are toric. We expect that we can obtain
generic toric Tri-Sasakian manifolds in this way, as in the case of
toric Kahler manifolds.

A single quaternion $q$, which is equivalent to a flat $R^4$, is given
by $q=q^4 I_2+i\vec\sigma\cdot\vec q$ with four real numbers $(q^4,\vec q)$.
It is also useful to introduce two complex variables $u$ and $v$ to write
\be
q=\left(\ba{cc} u & v \\ -\bar v & \bar u \ea \right)\quad.
\ee
The flat metric is $ds^2={1\over 2}{\rm tr}\left(dq d\bar
q\right)=dud\bar u+dv d\bar v$, and the triplet of Kahler forms of
hyper-Kahler structure is given by $\vec\omega\cdot\vec\sigma={1\over
2}dq\wedge d\bar q$, or more explicitly
\bear
\omega^3 &=&-{i\over 2}(du\wedge d\bar u+dv\wedge d\bar v)\quad,\nonumber\\
(\omega^1-i\omega^2) &= & i(du \wedge dv)\quad.\label{forms}
\eear
Under a $U(1)$-action of charge $Q$ defined by $q\to q\cdot e^{i
Q\sigma_3\xi }$, where $\xi$ is  an angle variable, we easily see that
the triplet of Kahler forms $\vec\omega$ is invariant, and we will use
these actions later to perform hyper-Kahler quotients.
In components, this corresponds to $u\to u\cdot e^{iQ\xi}$ and $v\to v\cdot e^{-iQ\xi}$.
Under the $SU(2)_R$ symmetry given by
\be
q\to \exp\left(-{i\over2}\vec\epsilon\cdot\vec\sigma\right)q\quad,
\ee
the Kahler forms transform as a triplet.

We will start from a $(n+r)$-dimensional flat quaternion space $(q_a), a=1,\ldots,(n+r)$, or flat
$R^{4(n+r)}$,
and consider $G=U(1)^r$ action under which $q_a$ has integer charges $Q_a^i, i=1,\ldots,r$.
After performing hyper-Kahler quotient by $G=U(1)^r$, we obtain a
$(4n)$-dimensional hyper-Kahler cone and its $(4n-1)$-dimensional
Tri-Sasakian cross-section, which is labeled by the charges $Q_a^i$.
We will calculate its volume as well as the volumes of its co-dimension
1 supersymmetric cycles via the equivariant localization technique in
the previous section.
The results are simple integration formulae in terms of $Q_a^i$, and in
many cases they can be explicitly calculated into campact expressions, as we will
see in several examples. 

It is worth commenting a difference between our case of Tri-Sasakian
manifolds and the case of Sasaki-Einstein manifolds from Calabi-Yau
cones.
In the case of toric Calabi-Yau cones or Sasaki-Einstein manifolds obtained
from  Kahler quotient of flat spaces, the induced metric on the quotient
space from the
ambient flat space is not in general Calabi-Yau, although the Kahler
class coincides. The actual Ricci-flat Kahler metric must be found in
other ways, and this is all about recent development of Z-minimization \cite{Martelli:2005tp,Martelli:2006yb}.
In the dual $\CN=1$ SCFT side, this corresponds to non-rigidity of
$U(1)_R$ symmetry under RG-flow, and the fixing of the  $U(1)_R$ symmetry by
a-maximization \cite{Intriligator:2003jj} at
IR fixed point is equivalent to finding Ricci-flat Kahler metric in the
gravity side. On the contrary, hyper-Kahler structure that we are
interested in for Tri-Sasakian manifolds is more rigid, essentially
because of its non-abelian $SU(2)_R$ structure, and the induced metric
from the flat ambient space is already hyper-Kahler that we  need.
This is a basic reason behind  our ability to calculate the volumes
via equivariant localization. In 3D $\CN=3$ SCFT side, this is related to
the fact that $SU(2)_R$ symmetry is rigid under RG-flow due to its
non-abelian nature.

{\it Toric Tri-Sasakian Manifolds}

The $G=U(1)^r$-action on $(n+r)$-dimensional quaternion space
$H^{(n+r)}$, $(q_a), a=1,\ldots,(n+r)$ with charges $Q_a^i,
i=1,\ldots,r$, is given by
\be
q_a\to q_a\cdot e^{iQ_a^i\sigma^3 \xi_i}\quad,
\ee
or in terms of complex coordinates  $u_a\to u_a \cdot e^{iQ_a^i\xi_i}$
and $v_a\to v_a \cdot e^{-iQ_a^i\xi_i}$.
The generating vector fields of $G$ commuting with each other are
\be
V^i={\partial\over\partial\xi_i}=i\sum_a
Q_a^i\left(u_a{\partial\over\partial u_a}-\bar
u_a{\partial\over\partial\bar u_a}-v_a{\partial\over\partial v_a}+\bar
v_a{\partial\over\partial\bar v_a}\right)\quad,
\ee
and with (\ref{forms}), the corresponding moment maps defined by
$i_{V^i}\vec\omega=Q\vec\mu_i$ are easily calculated to be
\be
\mu_i^3={1\over 2}\sum_a Q_a^i\left(|u_a|^2-|v_a|^2\right)\quad,\quad
(\mu^1_i-i\mu^2_i)=-\sum_a Q_a^i u_a v_a\quad.
\ee
It is also easy to find a $U(1)_R\subset SU(2)_R$ symmetry with the
properties $\CL_R \omega^3=0$,
$\CL_R(\omega^1-i\omega^2)=2i(\omega^1-i\omega^2)$, and
$\CL_R(\mu^1_i-i\mu^2_i)=R^\mu\partial_\mu(\mu^1_i-i\mu^2_i)=2i(\mu^1_i-i\mu^2_i)$
for all $i$,
to implement equivariant $\epsilon$-deformation that we discussed in the
previous section,
\be
R=i\sum_a \left(u_a{\partial\over\partial u_a}-\bar
u_a{\partial\over\partial\bar u_a}+v_a{\partial\over\partial v_a}-\bar
v_a{\partial\over\partial\bar v_a}\right)\quad.
\ee

The crucial fact that makes it possible to extract the volume of
the transverse Tri-Sasakian
section out of the $\epsilon$-deformed volume of the hyper-Kahler cone is
that
the function $H$ defined by $i_R \omega^3=Q H$ turns out to be simply
$H={1\over 2}\sum_a \left(|u_a|^2+|v_a|^2\right)={1\over 2}r^2$, where
$r$ is the radial coordinate of the hyper-Kahler cone. Therefore the
resulting expression (\ref{regvolume}) of the $\epsilon$-deformed volume of our hyper-Kahler cone
labeled by charges $Q_a^i$,
\be
{\rm Vol}_\epsilon\left(H^{(n+r)}////U(1)^r\right)={(4\pi\epsilon)^r \over (2\pi)^{3r}{\rm Vol}(U(1)^r)}
\int_{T[1]H^{(n+r)}\otimes\phi_3^i}\,\,e^{\omega+i\phi_3^i\mu^3_i(x)-\epsilon H}\quad,\label{rhs}
\ee
must be equal to the regularized volume of the cone $ds^2_{4n}=dr^2+r^2
ds^2_{4n-1}$ with a damping factor $e^{-{\epsilon\over 2}r^2}$, which is nothing
but
\be
{\rm Vol}_\epsilon\left(H^{(n+r)}////U(1)^r\right)={2^{2n-1}\Gamma(2n)\over
\epsilon^{2n}}\cdot {\rm Vol}\left(X_{4n-1}\right)\quad,\label{regu}
\ee
with the normalized volume of our Tri-Sasakian space $X_{4n-1}$.
The integration over $T[1]H^{(n+r)}$ in the right-hand side of (\ref{rhs})
is simple Gaussian integration for both bosonic and fermionic variables,
and we can caculate it easily. This is the pay-off that we receive for
all the previous technicalities of reformulating things in terms of the flat
ambient space where calculations become tractible. What remains is an
integration over $\phi_3^i\equiv \phi^i, i=1,\ldots,r$, and the resulting
expression for the right-hand side of (\ref{rhs}) is
\be
{\rm Vol}_\epsilon\left(H^{(n+r)}////U(1)^r\right)
={2^r (2\pi)^{2n}\over \epsilon^{2n}{\rm Vol}\left(U(1)^r\right)}\int
\prod_{i=1}^r d\phi^i \,\,\prod_{a=1}^{n+r} {1\over 1+\left(\sum_{i=1}^r Q_a^i
\phi^i\right)^2}\quad,
\ee
and by comparing with (\ref{regu}), we finally obtain the result for the
normalized volume of our Tri-Sasakian manifold labeled by $Q_a^i$
\be
 {\rm Vol}\left(X_{4n-1}\right)
={2^{r+1} \pi^{2n}\over \Gamma(2n){\rm Vol}\left(U(1)^r\right)}\int
\prod_{i=1}^r d\phi^i \,\,\prod_{a=1}^{n+r} {1\over 1+\left(\sum_{i=1}^r Q_a^i
\phi^i\right)^2}\quad.\label{result}
\ee
Note that ${\rm Vol}\left(U(1)^r\right)$ that is defined in the previous section is  the
coordinate volume of the $r$-dimensional torus defined by the angle
variables $\xi_i$.
That is, we identify $\xi_i\sim \xi_i+\eta_i$
whenever $\sum_{i=1}^r Q_a^i\eta_i\in 2\pi Z$ for all
$a=1,\ldots,(n+r)$, and  ${\rm Vol}\left(U(1)^r\right)$ is the volume of
an unit cell in the $r$-dimensional $\xi_i$ space. As it should be, our
result (\ref{result}) is invariant under an overall rescaling of $Q_a^i$ because of
this ${\rm Vol}\left(U(1)^r\right)$ factor, and we can consider only
relatively prime set of charges $Q_a^i$.

{\it Examples}

Let us consider the simplest example with $n=r=1$, which is a
3-dimensional Tri-Sasakian manifold obtained by  $U(1)$ hyper-Kahler
quotient from $H^2$,
 labeled by a relatively prime pair
of integer charges $(Q_1,Q_2)$. Explicitly, in terms of complex
variables of two flat quaternions $q_1\sim(u_1,v_1)$ and $q_2\sim(u_2,v_2)$,
the moment map equations are
\be
Q_1\left(|u_1|^2-|v_1|^2\right)+Q_2\left(|u_2|^2-|v_2|^2\right)=0\quad,\quad
Q_1 u_1 v_1 +Q_2 u_2 v_2 =0\quad,
\ee
with the identification
$\left(u_1,v_1,u_2,v_2\right)\sim\left(u_1e^{iQ_1\xi},v_1e^{-iQ_1\xi}, u_2e^{iQ_2\xi},v_2e^{-iQ_2\xi}\right)$.
The first equation in the above with this $U(1)$-quotient is a usual
Kahler quotient of $C^4$ with charges $(Q_1,-Q_1,Q_2.-Q_2)$ which is
also equivalent to a holomorphic quotient of $C^4$ by a $C^*$ action
\be
(u_1,v_1,u_2,v_2)\sim
\left(\lambda^{Q_1}u_1,\lambda^{-Q_1}v_1,\lambda^{Q_2}u_2,\lambda^{-Q_2}v_2\right)
\quad,\quad \lambda\in C^*\quad.
\ee
Introducing $Z_1=u_1 v_1$, $Z_2=u_2 v_2$, $Z_3=u_1^{Q_2}v_2^{Q_1}$ and
$Z_4=u_2^{Q_1}v_1^{Q_2}$ satisfying an algebraic equation
$Z_1^{Q_2}Z_2^{Q_1}=Z_3 Z_4$ that defines a hypersurface in $C^4$, it is
not difficult to show that the above map from $(u_i,v_i)$ to $Z_i$ is a bijection from  $C^4/C^*$
to the hypersurface $Z_1^{Q_2}Z_2^{Q_1}=Z_3 Z_4$ in $C^4$.
The fact that $(Q_1.Q_2)$ are relatively prime integers is used in
showing the above equivalence.
In terms of $Z_i$, the remaining equation $Q_1u_1v_1+Q_2u_2v_2=0$ becomes $Q_1
Z_1+Q_2 Z_2=0$, and therefore our hyper-Kahler cone is identified as  an algebraic variety
in $C^4$ defined by two equations
\be
Z_1^{Q_2}Z_2^{Q_1}=Z_3 Z_4\quad,\quad
Q_1Z_1+Q_2Z_2=0\quad,
\ee
and after solving for $Z_2=-{Q_1\over Q_2}Z_1$ and inserting into the
first equation, it becomes up to constant rescaling, a variety in $C^3$
described by $Z_1^{Q_1+Q_2}=Z_3 Z_4$. This is nothing but
an ALE space with $A_{(Q_1+Q_2-1)}$-singularity or $C^2/Z_{(Q_1+Q_2)}$ embedded into $C^3$, which
is indeed a hyper-Kahler cone. The corresponding Tri-Sasakian manifold
is then simply $S^3/Z_{(Q_1+Q_2)}$ or the Lens space $L_{(Q_1+Q_2)}$,
whose normalized volume must be
\be
{{\rm Vol}\left(S^3\right)\over \left(Q_1+Q_2\right)}={2\pi^2 \over
\left(Q_1+Q_2\right)}\quad.
\ee
To check that our formula (\ref{result}) indeed reproduces this, note
that  ${\rm Vol}\left(U(1)\right)$ is
the minimal number $\eta$ with $Q_1\eta, Q_2\eta \in 2\pi Z$, which is
simply $2\pi$ for relatively prime $(Q_1,Q_2)$.
We then have
\be
{\rm Vol}\left(X_3\right)=2\pi\int d\phi \,{1\over
(1+Q_1^2\phi^2)(1+Q_2^2\phi^2)}=2\pi\cdot{\pi\over\left(Q_1+Q_2\right)}={2\pi^2\over
\left(Q_1+Q_2\right)}\quad,
\ee
which agrees with the previous result.

In the context of M-theory in $AdS_4\times X_7$ that is dual to a 3D
$\CN=3$ SCFT, the next  example is $n=2$ with $r=1$ \cite{Lee:2006ys}. The resulting
7-dimensional space is labeled by three integer charges $(Q_1,Q_2,Q_3)$ being
relatively prime. With
\be
{\rm Vol}\left(U(1)\right)=(2\pi)l.c.m.\left(1\over
Q_i\right)={2\pi\over Q_1Q_2Q_3} l.c.m. (Q_1Q_2,Q_2Q_3,Q_3Q_1)=2\pi\quad,
\ee
where $l.c.m.$ stands for least common multiple, the right-hand side of
(\ref{result})
is readily calculated to be
\be
{\rm Vol}\left(X_7(Q_1,Q_2,Q_3)\right)={\pi^4\over 3}
{(Q_1Q_2+Q_2Q_3+Q_3Q_1)\over
(Q_1+Q_2)(Q_2+Q_3)(Q_3+Q_1)}\quad,
\ee
which includes the unique homogeneous 7-dimensional Tri-Sasakian manifold
$N(1,1)=SU(3)/U(1)$ as a special case of $Q_1=Q_2=Q_3=1$ with volume $\pi^4\over 8$.

{\it Codimension 1 Cycles}

In M theory on $AdS_4\times X_7$, an M5-brane wrapping a supersymmetric
codimension 1 cycle $\Sigma_5$ in $X_7$ looks like a very heavy point-like excitation in
$AdS_4$ spacetime. Its mass in $AdS_4$ is proportional to the volume of
the cycle, and by the standard AdS/CFT relation between mass in AdS and
the conformal dimension $\Delta$ of the corresponding operator in the dual CFT,
we have
\be
\Delta={\pi N\over 6}{{\rm Vol}\left(\Sigma_5\right)\over {\rm
Vol}\left(X_7\right)}\quad,
\ee
where $N$ is the number of background M2-brane charge in $AdS_4\times X_7$ \cite{Gubser:1998fp}. We are
therefore interested in volumes of codimension 1 cycles.

We will calculate volumes of supersymmetric codimension 1 cycles
defined by a holomorphic constraint $u_a=0$ or $v_a=0$ for some $a$, for
generic toric Tri-Sasakian manifolds labeled by $Q_a^i$.
Since the result turns out to be independent of $a$, we will simply
consider the cycle defined by $u_1=0$.
In the flat ambient space $H^{(n+r)}$ before performing hyper-Kahler
quotient, the hypersurface $u_1=0$ is Poincare dual to the 2-form
\be
\Gamma_2=\delta(u_1)\delta(\bar u_1)\psi^{u_1}\psi^{\bar u_1}\quad,
\ee
with $Q\Gamma_2=0$.
Though it is divergent, the symplectic volume of the hypersurface is represented formally
by the expectation value of the dual 2-form $\Gamma_2$ in the previous superspace formalism of
the symplectic volume of $H^{(n+r)}$;
\be
{\rm Vol}(\Sigma_{u_1=0})=\langle \Gamma_2 \rangle =\int_{T[1]H^{(n+r)}}\,\Gamma_2 \,\,e^S\quad,
\ee
with $S=\omega$. The delta functions in $\Gamma_2$ restrict the integration onto
the hypersurface $u_1=0$ and the fermionic factor takes care of reduction of the degree
of the volume form on $u_1=0$. The fact $Q\Gamma_2=0$ is interpreted as  $\Gamma_2$
being a nice observable in the $Q$-cohomology.

Getting back to our hyper-Kahler cone obtained by a hyper-Kahler quotient of $G=U(1)^r$ with charges $Q_a^i$,
we previously described the quotient space  in terms of the ambient superspace
$T[1]H^{(n+r)}\otimes \vec\phi^i\otimes \chi_a^i$, $i=1,\ldots,r, a=1,2$,
with a fermionic symmetry $\tilde Q$ in (\ref{tildeQ}),
and the action $S=\omega+i\vec\phi^i\cdot\vec\mu_i(x)+\chi_a^i Q\mu^a_i (x)$.
According to the spirit of equivariant cohomology, the usual $Q$-cohomology of the quotient space
is equivalent to the $\tilde Q$-cohomology in the ambient space. Moreover, expectation values
of cohomology elements are also expected to agree with each other.
The hypersurface defined by $u_1=0$ in the quotient
space should be described by some Poincare dual 2-form, and to describe it in terms of
the ambient superspace, we need to find a suitable generalization of $\Gamma_2$ satisfying $\tilde Q\tilde \Gamma_2=0$.
Fortunately, due to delta function factor in $\Gamma_2$, we simply have $\tilde \Gamma_2=\Gamma_2$.

Note that the hypersurface $u_1=0$ in our $4n$-dimensional hyper-Kahler cone is
another cone with dimension $4n-2$. As our Tri-Sasakian manifold $X_{4n-1}$
is the unit radius section of the hyper-Kahler cone,
its codimension 1 cycle $\Sigma_{4n-3}$ that we are interested in is the unit radius section of the hypersurface $u_1=0$.
We have previously introduced an $\epsilon$-deformation using $U(1)_R \subset SU(2)_R$ symmetry,
which is nothing but a damping factor $-{\epsilon\over2} r^2$,
where $r$ is the radial coordinate of the cone. From the resulting $\epsilon$-regularized
volume of the cone, we were able to extract the volume of the unit radius section of the cone.
After we deform the fermionic symmetry from $\tilde Q$ to $\tilde Q_\epsilon$ in (\ref{deformedQ})
in addition to $S_\epsilon=S-\epsilon H = S-{\epsilon\over 2}r^2$, the $\epsilon$-regularized
volume of the hypersurface $u_1=0$ is still expected to be an expectation value
of an observable $\tilde\Gamma_2$ with $\tilde Q_\epsilon \tilde\Gamma_2=0$.
Again due to the delta function factor, we easily find that $\Gamma_2$ satisfies $\tilde Q_\epsilon \Gamma_2=0$.

In summary, the regularized volume of the cone $u_1=0$ with dimension $4n-2$ inside our quotient space
is simply obtained by inserting $\Gamma_2$ in the partition function (\ref{rhs}),
\be
\langle \Gamma_2 \rangle_\epsilon={(4\pi\epsilon)^r \over (2\pi)^{3r}{\rm Vol}(U(1)^r)}
\int_{T[1]H^{(n+r)}\otimes\phi_3^i}\,\,\Gamma_2\,\,e^{\omega+i\phi_3^i\mu^3_i(x)-\epsilon H}\quad.\label{cone}
\ee
Since the regularization is a simple factor $-{\epsilon\over 2}r^2$, the above must be equal to
\be
{2^{2n-2}\Gamma(2n-1)\over \epsilon^{2n-1}}\cdot {\rm Vol}\left(\Sigma_{4n-3}\right)\quad,
\ee
where $\Sigma_{4n-3}$ is the unit radius section, which is our codimension 1 cycle inside $X_{4n-1}$.
The Gaussian integration is readily calculated as before to arrive at
\bear
{\rm Vol}\left(\Sigma_{4n-3}\right)
&=&{2^{r+1} \pi^{2n-1}\over \Gamma(2n-1){\rm Vol}\left(U(1)^r\right)}\int
\prod_{i=1}^r d\phi^i \,\,{1 \over \left(1+i\sum_{i=1}^r
Q_1^i\phi_i\right)}\cdot\prod_{a=2}^{n+r} {1\over 1+\left(\sum_{i=1}^r Q_a^i
\phi^i\right)^2}\nonumber \\
&=&{2^{r+1} \pi^{2n-1}\over \Gamma(2n-1){\rm Vol}\left(U(1)^r\right)}\int
\prod_{i=1}^r d\phi^i \,\,{1-i \sum_{i=1}^r
Q_1^i\phi_i\over 1+\left(\sum_{i=1}^r
Q_1^i\phi_i\right)^2}\cdot\prod_{a=2}^{n+r} {1\over 1+\left(\sum_{i=1}^r Q_a^i
\phi^i\right)^2}\nonumber \\
&=&{2^{r+1} \pi^{2n-1}\over \Gamma(2n-1){\rm Vol}\left(U(1)^r\right)}\int
\prod_{i=1}^r d\phi^i \,\,{1\over 1+\left(\sum_{i=1}^r
Q_1^i\phi_i\right)^2}\cdot\prod_{a=2}^{n+r} {1\over 1+\left(\sum_{i=1}^r Q_a^i
\phi^i\right)^2}\nonumber \\
&=&{2^{r+1} \pi^{2n-1}\over \Gamma(2n-1){\rm Vol}\left(U(1)^r\right)}\int
\prod_{i=1}^r d\phi^i \,\,\prod_{a=1}^{n+r} {1\over 1+\left(\sum_{i=1}^r Q_a^i
\phi^i\right)^2}\quad,
\label{cycleresult}
\eear
where in the second line the imaginary part vanishes under the integration over $\phi^i$.

An interesting fact is that the above looks very similar to the volume expression of $X_{4n-1}$
in (\ref{result}), in fact, the complicated integration over $\phi^i$ is identical.
Their ratio is
\be
{{\rm Vol}\left(\Sigma_{4n-3}\right)\over {\rm Vol}\left(X_{4n-1}\right)} ={1\over \pi}{\Gamma(2n)\over
\Gamma(2n-1)}={(2n-1) \over\pi}\quad,
\ee
which depends only on the dimension $(4n-1)$ of the Tri-Sasakian space
without regard to the quotient group $G=U(1)^r$.
It would be interesting
to find an underlying mathematical reason behind this universality.
Its consequence in AdS/CFT correspondence
of M-theory in $AdS_4\times X_7$ is that the conformal dimension of chiral primary
baryonic operators in $\CN=3$ SCFT
is always $\Delta={N\over 2}$.


\section{(2+1)D $\CN=3$ Field Theories}

{\it Some facts about (2+1)D field theories}

Let us first recall some unusual subtleties in (2+1)D field theories,
as we will need those in discussing $\CN=3$ supersymmetric theory.
For a massive excitation, we can go to its rest frame and its spin $s$ is
defined as the charge of $U(1)=SO(2)$ spatial rotation of $R^2$. Note that the
signature of $s$ is meaningful as it is impossible to flip its sign using Lorentz transformation,
contrary to (3+1)D case. In fact, $s$ is invariant under CPT and particles and anti-particles have
the same spin $s$,
\be
s \longrightarrow_P -s \longrightarrow_T s \longrightarrow_C s\quad.
\ee
Since $s$ flips its sign under parity $P$, a theory with particles of spin $s$ without particles of $-s$ breaks
parity.
Because we can exchange two identical excitations on the spatial $R^2$ plane, there still exists
the concepts of statistics, and the usual spin-statistics theorem holds true in (2+1)D.
A statistical phase under exchange of two identical particles must form a representation
of $\pi_1(RP^1)=\pi_1(S^1)=Z$ and can take an arbitrary $U(1)$ phase, though we will
only concern about bosons and fermions in the usual sense.

An example of massive, parity breaking theories is the Maxwell-Chern-Simons theory \cite{Deser:1981wh},
\be
\CL=-{1\over 4} F_{\mu\nu}F^{\mu\nu}+{\kappa\over 4}\epsilon_{\mu\nu\lambda}A_{\mu}F_{\nu\lambda}
-{\xi\over 2}(\partial_\mu A^\mu)^2\quad,
\ee
which describes spin $s={\kappa\over|\kappa|}$ excitations with $m^2=\kappa^2$ upon quantization.
To see this briefly, the operator equation of motion in $\xi=1$ gauge is
\be
\left(\partial^2 \eta^{\mu\nu}-\kappa \epsilon^{\mu\nu\lambda}\partial_\lambda\right)A_{\nu}=0\quad,
\ee
and expanding in terms of creation/annihilation operators $a^\dagger(\vec p)$ and $a(\vec p)$,
\be
A_\mu(x)=\int{d^2\vec p\over(2\pi)^2\sqrt{2p_0}}\,\left[\varepsilon_\mu(\vec p) e^{-ip\cdot x}
a(\vec p)+{\rm h.c.}\right]\quad,\ee
its polarization $\varepsilon_\mu(\vec p)$ for particle excitations satisfies
\be
\left(p^2 \eta^{\mu\nu}-i\kappa\epsilon^{\mu\nu\lambda}p_\lambda\right)\varepsilon_\nu=0\quad,
\ee
whose non-trivial solution exists uniquely when $p^2=-\kappa^2$. To determine the spin,
we go to the rest frame of particle excitation $p_\mu=(|\kappa|,\vec0)$ where the polarization becomes
\be
\varepsilon_\mu={1\over \sqrt{2}}\left(\ba{c}0\\1\\i{\kappa\over|\kappa|}\ea\right)\quad,
\ee
which has the charge $s={\kappa\over|\kappa|}$ under spatial $SO(2)$ rotation.
Note that there is no additional anti-particle excitations in the theory.

Another example of parity breaking theories is a massive Majorana fermion in (2+1)D.
In (2+1)-dimension, two-components Majorana representation is possible with
\be
\gamma^0=i\sigma^2\quad,\quad \gamma^1=\sigma^1\quad,\quad\gamma^2=\sigma^3\quad.
\ee
A Majorana spinor $\psi=(\psi_1,\psi_2)^T$ transforms as
$(\psi_1,\psi_2)^T\to(e^{\alpha\over 2}\psi_1,e^{-{\alpha\over 2}}\psi_2)^T$ under a boost, and
$(\psi_1,\psi_2)^T\to (\cos({\phi\over 2})\psi_1 -\sin({\phi\over 2})\psi_2,\sin({\phi\over 2})\psi_1 +\cos({\phi\over 2})\psi_2)^T$
under a rotation of angle $\phi$, which are consistent with the reality of $\psi$.
The massive Lagrangian is
\be
\CL=i\bar\psi \gamma^\mu\partial_\mu \psi+i m \bar\psi \psi\quad,
\ee
whose equation of motion $i\gamma^\mu\partial_\mu \psi+im\psi=0$  in components becomes
\be
(\partial_2+m)\psi_1 +(\partial_0+\partial_1)\psi_2=0\quad,\quad
(-\partial_0+\partial_1)\psi_1+(-\partial_2+m)\psi_2=0\quad.
\ee
The non-trivial solution exists only when $p^2=-m^2$ as expected, with the form
\be
\psi=\left(\ba{c} 1\\-\left({p_2+im\over p_0+p_1}\right)\ea\right) e^{-ip\cdot x}a(\vec p)+{\rm h.c.}\quad,
\ee
for the mode with momentum $\vec p$. In the rest frame $p_\mu=(|m|,\vec 0)$,
the wave function looks like $(1, -i{m\over|m|})^T$, which has the charge $s=-{1\over 2}{m\over|m|}$
under spatial rotation of angle $\phi$. Therefore, the spin depends on the sign of the mass term,
and indeed the mass term in the Lagrangian is odd under parity. The usual parity conserving Dirac mass term
for a Dirac spinor can be considered as two Majorana fermions with opposite sign of mass terms.

If we take massless limit, the relevant concept would be helicity rather than spin
as there doesn't exist a rest frame at all, but
since there is no little group in (2+1)-dimension, helicity does not exist either.
This can be understood in an example of
the duality between $U(1)$ gauge theory and a periodic real scalar field theory in (2+1)D.
The statistics whether excitations are bosons or fermions remains meaningful
in the massless theory from (anti)commutator algebra of operators.
This means that
the Hilbert space allows an operator $(-1)^F$ which anti-commutes
with all fermionic operators.

{\it A warm-up with $\CN=1$}

The basic unit of supercharges in (2+1)D is a Majorana spinor with two real components.
From the Lorentz transformation of the $\CN=1$ supercharge $Q_\alpha=(Q_1,Q_2)^T$,
we see that $Q_+={1\over\sqrt{2}}(Q_1+iQ_2)$ has spin $1\over 2$, and $Q_-=(Q_+)^\dagger={1\over\sqrt{2}}(Q_1-iQ_2)$ has spin $-{1\over 2}$.
Since $(p_1\pm ip_2)$ have spin $\pm 1$ and $p_0=E$ has spin $0$, we expect
\be
\{Q_+, Q_-\}=E\quad,\quad\{Q_+,Q_+\}=(p_1+ip_2)\quad,\quad \{Q_-,Q_-\}=(p_1-ip_2)\quad,
\ee
which can be written in a covariant way as
\be
\{Q_\alpha,Q_\beta\}=-\left(\gamma^0\gamma^\mu p_\mu\right)_{\alpha\beta} \quad.
\ee

For a massless excitation of a $\CN=1$ theory, we go to the frame with $E=p_1$ and $p_2=0$, where we have
\be
\{Q_1,Q_1\}=E\quad,\quad\{Q_2,Q_2\}=\{Q_1,Q_2\}=0\quad.
\ee
The minimal representation is a pair of bosonic/fermionic excitations $|b\rangle$,$|f\rangle$
with $|f\rangle=Q_1|b\rangle$ and $|b\rangle=Q_1|f\rangle$.
This is easily realized by the minimal super Yang-Mills theory of gauge field and Majorana gaugino,
\be
\CL=-{1\over 4}F_{\mu\nu}F^{\mu\nu}+i\bar\lambda \gamma^\mu\partial_\mu\lambda\quad,
\ee
which is invariant under $\delta A_\mu=i\bar\epsilon \gamma_\mu\lambda=-i\bar\lambda\gamma_\mu\epsilon$
and $\delta \lambda=-{1\over 4}F_{\mu\nu}\gamma^{\mu\nu}\epsilon$.
For $U(1)$ theory, it is equivalent to a theory of real scalar field and a Majorana fermion using abelian duality,
\be
\CL={1\over 2}\partial_\mu\phi \partial^\mu\phi +i\bar\lambda \gamma^\mu\partial_\mu\lambda\quad,
\ee
with a supersymmetry
$\delta\phi=i\bar\epsilon\lambda$ and $\delta\lambda=-{1\over 2}\partial_\mu\phi\gamma^\mu\epsilon$.
The two supersymmetry transformations agree with each other under the duality transformation.

This duality breaks down under mass perturbations.
For massive excitations, we go to the rest frame $p_\mu=(E,0,0)$ where
\be
\{Q_+,Q_-\}=E\quad,\quad \{Q_+,Q_+\}=\{Q_-,Q_-\}=0\quad.
\ee
Starting from a state $|s\rangle$ with spin $s$, the state $Q_+|s\rangle=|s+{1\over 2}\rangle$ has spin $s+{1\over 2}$.
Up to parity transformation, there are only two cases for field theory; for $s=0$ we have a massive theory of real scalar and a
Majorana fermion, and for $s={1\over 2}$ we have a massive version of minimal super Yang-Mills theory.
For Yang-Mills theory, there are two ways of introducing masses. The one is through a Higgs mechanism
which does not break parity. This means that elementary bosonic excitations consist of both signs of spin $s=\pm 1$.
Since we then have two bosonic degrees of freedom, we must have two Majorana fermions with a parity conserving
Dirac mass term, in other words, two Majorana mass terms with opposite signs.
Note that we have to double the degrees of freedom to preserve parity.
To avoid doubling of excitations, we have to break parity.
The Chern-Simons term for gauge field and a Majorana mass term for gaugino that we have discussed before
do the job. The supersymmetry can be easily checked with these terms.
The massive theory of a real scalar field and a Majorana fermion is also simply obtained by adding usual mass terms
to the massless Lagrangian,
and it is physically different from the massive super Yang-Mills theory, although it also breaks parity.

{\it Massive $\CN=3$ theory }

We first discuss massive $\CN=3$ theory.
In the rest frame of an excitation, the supersymmetry algebra becomes
\be
\{Q^I_+,Q_-^J\}=E\,\delta^{IJ}\quad, I,J=1,2,3\quad,
\ee
with others vanish\footnote{We are neglecting central charges.}.
Starting from a state $|s\rangle$ of spin $s$ with $Q^I_-|s\rangle=0$, the resulting multiplet
is shown in the table below, where $SO(3)_R$ is the R-symmetry of three Majorana supercharges.
\begin{table}[h]
\begin{tabular}{|c|c|c|c|c|}
\hline
   &  $  |s\rangle$ & $Q_+^I |s\rangle$    &   $Q_+^I Q_+^J |s\rangle $ & $Q_+^1 Q_+^2 Q_+^3 |s\rangle$    \\
\hline
{\rm spin} &      $s$ & $s+{1\over 2}$ & $s+1$  & $s+{3\over 2}  $       \\
\hline
$SO(3)_R$  & $1$  & $3$  & $3$  & $1$ \\
\hline
\end{tabular}
\end{table}
Up to parity, the unique field theory multiplet is given by $s=-{1\over 2}$ for which
we have one spin 1 state, three spin $1\over 2$ states, three spin 0 states and one spin $-{1\over 2}$ state.
The field theory therefore has a gauge field, four Majorana fermions and three real scalars.
Incidentally the field content is identical to the $\CN=4$ massless vector multiplet, which
is obtained from a dimensional reduction of 6D $\CN=(1,0)$ vector multiplet.
This indicates a possibility that the massive $\CN=3$ theory may be obtained
by a mass perturbation to the massless $\CN=4$ super Yang-Mills theory.
Since we should not double the elementary excitations while giving masses to them,
the resulting theory will break parity.
The only known way to achieve this is to introduce
a Chern-Simons term for massive spin 1 particles
and Majorana mass terms for three spin $1\over 2$ fermions and single spin $-{1\over 2}$ fermion with opposite
sign of coefficient, and the usual mass terms for three scalars.
In an interacting theory, we need to add necessary coupling terms consistent with $\CN=3$.
This has been obtained in Ref.\cite{Kao:1992ig,Kao:1993gs} for non-abelian gauge theory,
\bear
\CL_{\CN=3} = \kappa\cdot{\rm tr}\left\{{1\over 2}\epsilon^{\mu\nu\rho}\left(A_\mu\partial_\nu A_\rho-i{2\over 3}
A_\mu A_\nu A_\rho\right)-i\bar\lambda_a \lambda_a +i\bar\chi \chi -\kappa C_a^2-{i\over 3}
\epsilon^{abc}C_a[C_b,C_c] \right\}\,,\nonumber
\eear
where $a,b,c=1,2,3$ are $SO(3)_R$ vector indices. Note that the signs of Majorana mass terms are consistent with
our expectation.
The Lagrangian of the massive $\CN=3$ theory is thus $\CL=\CL_{\CN=4}+\CL_{\CN=3}$.

In $\CN=4$, the R-symmetry is $SU(2)_1\times SU(2)_2$, one of which is the original R-symmetry
of 6D $\CN=(1,0)$ and the other comes from the spatial $R^3$ rotation of the reduced dimensions when we
dimensionally reduce to 3D. $(\lambda_a,\chi)$ is doublet under both $SU(2)$'s, while $C_a$ is a triplet
under the second $SU(2)_2$. The $\CN=3$ deformation preserves only the diagonal $SU(2)_D$ of $SU(2)_1\times SU(2)_2$,
which can be seen in the Majorana mass terms. The remaining $SU(2)_D$ is our R-symmetry $SO(3)_R$ of $\CN=3$.

{\it Massless $\CN=3$ theory}

For a massless excitation, the super-algebra in the frame $p_\mu=(E,E,0)$ is
\be
\{Q_1^I,Q_1^J\}=E\,\delta^{IJ}\quad,I,J=1,2,3\quad,
\ee
with $Q_2^I=0$.
Since $Q_1^I$'s are real, we seem to have an $E^3$-Clifford algebra, and the massless multiplet
must be a representation of it.
However, this is not a whole story. Since statistics is relevant in (2+1)D,
there must exist an operator $(-1)^F$ which anti-commutes with $Q_1^I$'s.
Including $(-1)^F$ in the algebra, we actually have an $E^4$-Clifford algebra, whose
minimal representation has dimension 4. Realizing the Clifford algebra by
\be
Q_1^{I=1}=\sigma^1\otimes\sigma^1\quad,\quad Q_1^{I=2}=\sigma^2\otimes\sigma^1\quad,\quad
Q_1^{I=3}=\sigma^3\otimes\sigma^1\quad,\quad (-1)^F=1\otimes \sigma^3\quad,\label{N=3}
\ee
the representation automatically allows another operator $Q_1^{I=4}=1\otimes\sigma^2$
which anti-commutes with $Q_1^I$, $I=1,2,3$ and $(-1)^F$.
This implies that a massless $\CN=3$ multiplet is automatically completed to
a massless $\CN=4$ multiplet.
Note however that this does not imply that a massless $\CN=3$ theory is also $\CN=4$,
because interactions do not necessarily preserve $\CN=4$ supersymmetry.
Since it is known that $\CN=3$ sigma model with a  hyper-Kahler manifold is automatically
$\CN=4$ \cite{Kapustin:1999ha}, a strict massless $\CN=3$ model should be something else.
One possibility is to have non-dynamical vector multiplets with only $\CN=3$ Chern-Simons terms
coupled with massless $\CN=4$ sigma model \cite{Kapustin:1999ha,Schwarz:2004yj}.

The dimension of minimal representation of $E^4$-Clifford algebra is 4, and
we expect two bosonic and two fermionic degrees of freedom.
Note that this is
half of the well-known $\CN=4$ vector/hyper multiplet from 6D $\CN=(1,0)$.
From the previous discussion of the massive $\CN=3$ theories, we also see that it would
be impossible to introduce mass perturbation without doubling the multiplet.
From the above explicit realization of the algebra (\ref{N=3}),
the $SO(3)_R$ R-symmetry generators are realized as
\be
R^I={\sigma^I\over 2}\otimes 1\quad,I=1,2,3\quad,
\ee
with $[R^I,Q_1^J]=i\epsilon^{IJK}Q_1^K$ and $[R^I,(-1)^F]=0$ as required.
This implies that two bosons as well as two fermions are both doublets under $SO(3)_R$.
If we would like to realize these in field theory, since a doublet of $SO(3)_R$ must be
complex, we would end up with four bosonic degrees of freedom, contradicting to the above.
This case is actually $\CN=4$ hyper-multiplet with two complex scalars and two Dirac fermions
which are doublets under $SU(2)_1\otimes SU(2)_2$ respectively. They both become doublets under the diagonal
$SO(3)_R$ in $\CN=3$. The massless $\CN=4$ vector multiplet is identified
by choosing somewhat complicated realization of $SO(3)_R$ generators, under which one real boson out of
two complex bosons is a singlet and the remaining three form a triplet.
These considerations indicate that in field theory level,
there is probably no minimal massless $\CN=3$ Lagrangian possible, although it does not exclude the possibility
of an intrinsic quantum theory realizing the above minimal representation.


\section{Dual $\CN=3$ SCFT Proposal}

In M-theory context, we consider N stack of M2 branes at the apex of a 8-dimensional
hyper-Kahler cone, whose 3D world-volume theory is $\CN=3$ SCFT at low energy.
Its supergravity solution has the near horizon geometry of $AdS_4\times X_7$ with
a background M2 charge flux N, where $X_7$ is the constant radius Tri-Sasakian section of
the original hyper-Kahler cone.
AdS/CFT correspondence conjectures a duality between these two descriptions.
The strongly coupled SCFT on M2 branes is mysterious.
From the analyses in the supergravity side, it has been found that the number of degrees of freedom
in 3D SCFT scales as $\sim N^{3\over 2}$, and its explanation is still missing.
Because this seems true without any regard to supersymmetry, its understanding
may lie in a new but general characteristic of strongly coupled 3D theories.

Although having supersymmetries does not help much to identify the nature of IR conformal field theory,
it can tell us some specific things that are protected by supersymmetries, such as scale dimensions
of chiral primaries, $SU(2)_R$ symmetry, etc.
The $\CN=3$ superconformal algebra dictates that for chiral primaries, $\Delta = j_R$
where $j_R$ is the spin number of the $SU(2)_R$ representation \cite{Minwalla:1997ka}.
Knowing charges under $SU(2)_R$ thus helps us to identify conformal dimensions.
An advantage over the case of $\CN=2$ where the R-symmetry $U(1)_R$ is not rigid under RG flow
is that for $\CN=3$, the non-abelian $SU(2)_R$ is invariant under RG flow, and can be fixed
in UV before flowing to a SCFT in IR.

This may allow us to propose a UV description in terms of quiver-type gauge theory,
which is supposed to flow into our $\CN=3$ SCFT in far infrared \cite{Fabbri:1999hw}.
This gauge theory in UV is not intended to give any dynamical information about the IR SCFT, such
as fundamental degrees of freedom or interactions between them.
Along the RG flow, the theory becomes strongly coupled and we no longer expect gauge fields
as our dynamical degrees of freedom.
The usefulness of the proposal for UV description, if it indeed flows to our SCFT in IR,
rests on the ability of identifying $SU(2)_R$ symmetry and the spectrum of chiral primary operators,
which should coincide with those in the IR SCFT fixed point due to rigidity under RG flow.

In attempts to propose a UV description of our $\CN=3$ SCFT that is dual to M-theory on $AdS_4\times X_7$,
there is one necessary condition we need to consider. In the original M2 brane picture,
there exists a Coulomb branch where M2 branes move apart from each other around the tip of the hyper-Kahler cone.
Our proposal then should have a moduli space of vacua corresponding to the Coulomb branch,
which is nothing but the N-symmetric product of our hyper-Kahler cone over the Tri-Sasakian manifold.
The simplest possibility as in Ref.(\cite{Klebanov:1998hh}) is to consider a non-abelian gauge theory whose vacuum conditions from D/F-terms
become a symmetric product of non-linear sigma model with target space given by our hyper-Kahler cone.
The symmetrization is a part of the original gauge symmetry.
The D/F-term equations from vector multiplets of $\CN=3$($\CN=4$) gauge theory are in fact moment-map
equations of a hyper-Kahler quotient, and this matches with the fact that our hyper-Kahler cone is also obtained
from a hyper-Kahler quotient.
Note that this is a simple guideline, and the actual proposal is not completely determined by this.

A massless $\CN=3$ multiplet is automatically $\CN=4$ multiplet as we see in the previous section.
To have strict $\CN=3$ supersymmetry, the only known way is to add a $\CN=3$ Chern-Simons term, which makes
vector multiplet massive with mass proportional to the coupling constant $e^2$.
We keep $\CN=4$ matter hyper-multiplets massless.
As we flow to IR, the gauge coupling constant blows up and the gauge fields become heavy and decouple.
An equivalent way of saying is that we can neglect kinetic terms for fields in the vector multiplets and they become
non-dynamical. However,
the Chern-Simons term does not disappear. Their effect in abelian case is to induce a non-zero
statistical phase for matter excitations to make them anyons \cite{Wilczek:1983cy}.
Therefore, a naive picture of physics in IR is to have a non-linear sigma model from hyper-multiplets
with target space obtained by D/F-term equations of vector multiplets, whose dynamics
is affected by non-dynamical vector multiplets with $\CN=3$ Chern-Simons terms.
However, the full IR dynamics will be more than the above, including loop excitations from M2 branes,
which presumably account for $ N^{3\over 2}$-scaling of degrees of freedom.

Having said the purposes and limitations of the UV proposals in terms of quiver gauge theories,
we give a simple proposal for the $\CN=3$ theory that corresponds to a 7D Tri-Sasakian manifold
obtained by an arbitrary $G=U(1)^r$ hyper-Kahler quotient
labelled by $Q_a^i$, with $i=1,\ldots,r$ and $a=1,\ldots,r+2$.
The gauge group is a $r$-copy of $U(N)\times U(N)$,
\be
\CG=\prod_{i=1}^r \left[U(N)\times U(N)\right]_i\quad.
\ee
Under the $i$'th group $\left[U(N)\times U(N)\right]_i$, the $(r+2)$ hyper-multiplets $U_a=(u_a, \bar v_a)$ are charged
as $\left(Sym^{Q_a^i},\bar{ Sym}^{Q_a^i}\right)$, where $Sym^Q$ stands for the $Q$-symmetric
representation of fundamental representation of $U(N)$. This is a naive generalization
of the previous proposal for the $r=1$ case \cite{Fabbri:1999hw,Lee:2006ys}.

The fundamental chiral primary fields $U_a=(u_a, \bar v_a)$ are doublet under $SU(2)_R$.
We can find baryonic gauge invariant chiral primary operators composed of N $U_a$'s for a fixed $a$
by contracting their gauge indices by $\epsilon^{1 2\cdots N}$-tensors. We need $2\sum_{i=1}^r Q^r_a$ number
of $\epsilon$-tensors to contract all gauge indices.
As this number is even, the resulting operator is totally symmetric under an exchange of any two $U_a$
components. Therefore, the baryonic operators for any fixed $a$ form a spin $j_R={N\over 2}$
representation under $SU(2)_R$ and their conformal dimension must be $\Delta={N\over 2}$ for all $a$.
The baryonic operators with given $a$ are mapped to a M5 brane wrapping the cycle $u_a=0$ or $v_a=0$.
Any other cycles that are obtained from this by $SU(2)_R$ action are all relevant.
We have to quantize the M5 brane moving along flat directions of supersymmetric cycles connecting $u_a=0$
and $v_a=0$ through $SU(2)_R$ orbit. As $u_a=0$ or $v_a=0$ is invariant under $U(1)_R \subset SU(2)_R$,
this orbit will be nothing but $SU(2)/U(1)=S^2$, and the M5 brane quantization
is identical to a problem of point-like particle on $S^2$. Due to the background M2 charge flux of N units,
this problem boils down to a charged particle on $S^2$ moving in a background monopole charge $N$.
The resulting spectrum agrees with the spin $j_R={N\over 2}$ multiplet of our baryon operators.
As we calculated in the previous sections, the conformal dimension from the geometry
side perfectly matches with this expectation.

As a final comment, there recently appeared a description of $\CN=2$ SCFT, arising from M2 branes
at the tip of toric $CY_4$ cones, in terms of crystals of M5 branes after T-duality \cite{Lee:2006hw}.
Since our $\CN=3$ SCFT's belong to $\CN=2$ SCFT, there must be a similar description.
It would also be interesting to think about the relation between our UV proposal of
quiver-type gauge theories and M5 crystals.

\vskip 1cm \centerline{\large \bf Acknowledgement} \vskip 0.5cm

The author is indebted to Kimyeong Lee, Sangmin Lee and Jae-Suk Park for valuable discussions.
He also appreciates kind invitations to Kobe University and Chuo University in Japan by Chong-Sa Lim
and Takeo Inami
respectively, where part of the work has been done.
This work was supported by the Korea Research Foundation Grant. (KRF-2005-070-c00030)

 \vfil

\end{document}